\documentclass[trackchanges, twocolumn]{aastex701}

\usepackage{hyperref}

\newcommand{\gfpm}{\texttt{GF~pastamarkers}}

\begin{document}

\title{New Paradigms in Pasta: Introducing \gfpm\ for Enhanced Inclusivity and Productivity}

\author{Julian Falcone}
\affiliation{Department of Physics and Astronomy, Georgia State University, 25 Park Place, Atlanta, GA 30303, USA}
\email{jfalcone2@gsu.edu}

\author{Nabanita Das}
\affiliation{Department of Physics and Astronomy, Georgia State University, 25 Park Place, Atlanta, GA 30303, USA}
\email{ndas5@gsu.edu}

\begin{abstract}

Informative data visualization methods are key to the clear and efficient communication of myriad forms of data. The PASTA Collaboration \citep{pasta24, PASTA25} have made substantial contributions to the field of data visualization through $\mathtt{pastamarkers}$, a Python-based package that utilizes various types of pasta as data markers to create engaging plots. This work introduces \texttt{GF pastamarkers}, an extension of \texttt{pastamarkers} that utilizes the tenuous structure of gluten free (GF) pasta to meet the needs of the GF population. The implementation of \gfpm\ employs an exponential crumbling factor ($CF$), which benefits authors by encouraging clearer and more concise scientific articles, thereby leading to more effective manuscripts and proposals.

\end{abstract}

\keywords{}

\section{Introduction} 

For as long as humans have lived on Earth, they have quantified the world around them and have endeavored to process those data through visual means. Thus, one can consider ancient cave paintings-- which exhibit graphical representations of human and animal populations for a given area-- to be among the oldest examples of data visualization, dating back almost 70,000 years from the present day \citep{Aubert18, Oktaviana26}. However, the term ``data visualization'' did not start to emerge in the human lexicon until the 2nd century AD \citep{Li20}. Common forms of early data visualization
included maps and trees, the latter of which typically comprised classifications of the natural world and of familial relationships \citep{Crampton01, Lima11}.

\begin{figure}[b]
\centering  
\includegraphics[width=\linewidth]{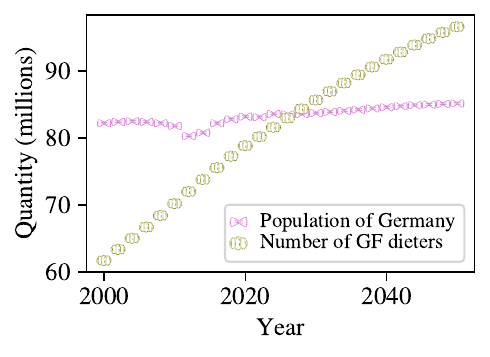}

\caption{A visualization of the estimated number of people around the world consuming a GF diet compared to projections of Germany's population up to the year 2050. The number of GF dieters is estimated to overtake the German population by 2027.}
\label{fig: population comparison}
\end{figure}

The advent of advanced astronomical instrumentation over the last century has ushered in an era of ``big data,'' requiring more complex tools to both analyze and visualize the data. Consequently, researchers are producing more plots in published works than ever before: as of 2013, the American Astronomical Society (AAS) journals published over 30,000 figures and 10,000 tables annually \citep{Biemesderfer13}. As of the writing of this manuscript in 2026, those numbers have surely risen considerably. As figures become easier to create and insert into published works, it becomes the chief responsibility of the authors to preclude superfluousness and ensure that only the most essential figures are included.

\subsection{Prelude to Gluten}

The etymology of ``gluten'' is derived from Latin in which the word directly translates to ``glue.'' Indeed, gluten is responsible for the elastic and rubbery texture of dough such that it can be easily stretched and manipulated without breaking \citep{Bailey02}.
As described as part of findings from the Third International Symposium on Sourdough \citep{Wieser07}, gluten contains hundreds of proteins that can be broken into roughly equal quantities of water-soluble gliadins and insoluble glutenins, and this equal distribution contributes to the physical properties of the dough.

It is estimated that $\sim$1\% of the worldwide population is affected by celiac disease, which is a serious autoimmune disease in which, upon the consumption of gluten, the body launches an aggressive attack on the small intestine \citep{Silano09}. Gluten generally refers to the proteins present in various grains, including wheat, rye, and barley \citep{Bailey02, Fasano12}. Foods that most commonly contain gluten include breads, cereals, pastries, and pastas. In Figure \ref{fig: population comparison}, we show a graphical representation of the number of people suffering from celiac disease compared to the historical and projected population of Germany.

Currently, the only proven treatment for celiac disease is a gluten-free (GF) diet \citep{Caio19}, which substitutes wheat flour with nut flour (such as almond or walnut) or other grains such as rice, corn, and millet. Demand for GF products has risen accordingly: sales in GF pasta reached \$1.52 billion in 2022 \citep{Pasini25} and are expected to grow annually at a compound rate of 7.6\% to 9.2\% \citep{Singla24}. Indeed, the popularity of GF pasta has expanded beyond those who are afflicted with celiac disease, and is sought after by those who believe in the health benefits of consuming GF food \citep{Winham22, Singla24}. Therefore, it becomes clear that the general population is searching for more ways to incorporate GF pasta into their daily lives.
In this work, we introduce \texttt{GF~pastamarkers}, which is an extension of the original \texttt{pastamarkers} package. By creating this GF option for the beloved markers, we enhance the inclusiveness of the field of astronomy and ensure that anybody regardless of dietary needs can utilize the whimsical textures of fusilli and cappalletti (among so many others) to create the plots of their dreams. The differences between the two, and the benefits of using \texttt{GF~pastamarkers} over the original version, are described in Section \ref{sec:method}.

\begin{figure}
\centering
\begin{minipage}{0.46\textwidth}
    \includegraphics[width=\linewidth]{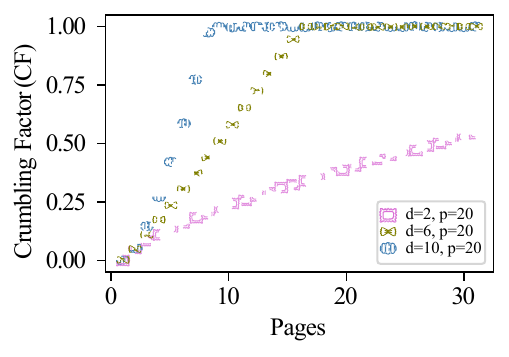}
\end{minipage} 
\begin{minipage}{0.46\textwidth}
    \includegraphics[width=\linewidth]{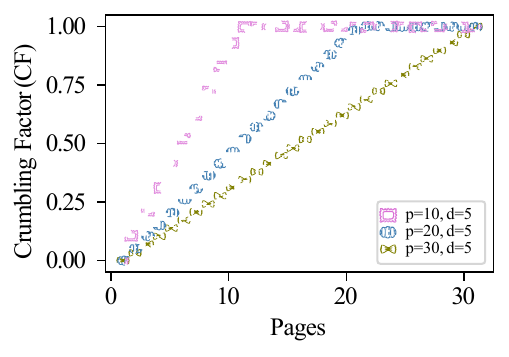}
\end{minipage}
\caption{A visualization of the $CF$ for different values of $p$ (top) and $d$ (bottom) as a function of paper length. The scale of crumbling ranges from $0-1$, where $CF=0$ presents the markers in their inherent shape and $CF=1$ presents completely crumbled markers. 
}
\label{fig: crumbling factor}
\end{figure}

\section{Justification \& Methodology} \label{sec:method}

Despite numerous breakthroughs in the formulation of GF products that strive to accurately replicate more traditional foods like pasta and bread, GF foods typically lack elasticity, and are therefore much more prone to breakage and crumbling \citep{Pasini25}. The primary improvement of \gfpm\ over \texttt{pastamarkers} is that it incorporates the characteristic crumbling tendency of GF foods in plots and graphs. Typically, this is an undesirable quality in foods because breakage and crumbling tend to negatively alter a food's texture. However, \gfpm\ utilizes crumbling in a novel way to help astronomers craft clearer, more concise manuscripts that will increase readability, lower publication costs, and improve chances of grant funding.

The benefits of \gfpm\ are exhibited in this manuscript, as is evident from the three figures included in this work. We define the crumble factor $(CF)$, which describes the extent to which the marker breaks down for a given page of a manuscript (denoted as $x$), as:
\begin{equation}
CF = \frac{1}{p}(x-1)^{\frac{d+5}{10}}.
\label{eq}
\end{equation}

Under this formalism, the $CF$ depends on two quasi-random factors. The variable $p$ represents the number of pages in a manuscript until $CF=1$, effectively rendering figures inscrutable. Thus, it is in authors' best interest to avoid reaching this limit by constructing shorter papers. The variable $d$ controls the curvature of the exponential trend. A few examples of $p$ and $d$, and the resulting impact on the $CF$, are exhibited in Figure \ref{fig: crumbling factor}.

The value for $p$ is determined by the number of visible penguins in the African Penguin Live Webcam at the Georgia Aquarium\footnote{https://georgiaaquarium.org/webcam/african-penguin-cam/} at the moment \gfpm\ is imported into the Jupyter notebook used to construct the relevant figures in authors' manuscripts. Based on the African Penguin population at this location (Georgia Aquarium, private communication), $1 < p < 50$. The value for $d$ is determined by the ones digit in the distance that the Voyager 1 spacecraft has traveled from Earth, in units of kilometers\footnote{https://science.nasa.gov/mission/voyager/where-are-voyager-1-and-voyager-2-now/}\footnote{https://theskylive.com/voyager1-info}, also at the moment \gfpm\ is imported into Jupyter notebook. In instances where this digit is zero, it gets rounded to 10. Thus, $1 < d < 10$.

The $p$ and $d$ values are stored in the metadata of the plot, and are not revealed until publication. Thus, authors must seriously consider the structure of their paper and consolidate as much as possible to minimize the extent of crumbling.

\subsection{Benefits of \gfpm}

\begin{figure}
    \centering
    \includegraphics[width=\linewidth]{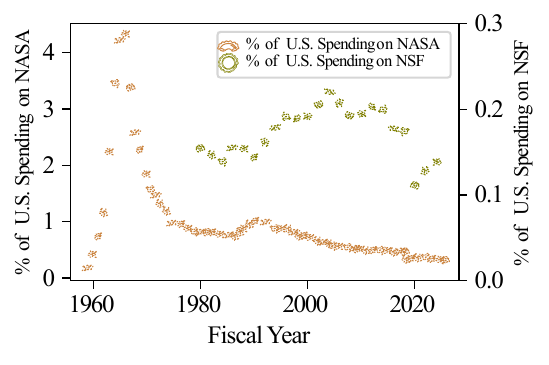}
    \caption{Percentage of United States budget spent on NASA and the NSF over time.
    The sources of the data in this plot are \citet{dreier_data} and the Office of Management and Budget and US Department of the Treasury, respectively.}
    \label{fig:budget}
\end{figure}

After reading the preceding section, authors may ask why it is worthwhile to utilize \gfpm\ in the first place. Why risk producing plots that disintegrate at a rate unbeknownst until it is too late? There are numerous reasons to use \gfpm\ that will yield benefits in both the short and long term. First, in recent decades the field of astronomy has taken extensive measures to create a more inclusive environment for all astronomers, and \gfpm\ contributes to that mission and responds to a growing need for GF accommodations across academia. Second, the beautiful designs of the pasta create an enticing appeal to the casual arXiv peruser, who may choose to read one paper over another due to aesthetic choices. However, this latter point could be said of the original \texttt{pastamarkers} package as well. Thus, we expand upon our final reason below to illustrate the specific benefit of \gfpm.

Since the success of the Apollo program in the 1960's, the federal budget's allocation for NASA has been steadily declining. As shown in Figure \ref{fig:budget}, in the 1970's $\sim$  0.7\% of the proposed budget was allocated for NASA missions, whereas today it has come down to a mere $\sim0.3-0.4$\% \citep{dreier}. According to the Office of Management and Budget and the US Department of the Treasury, the total budget allocated to the National Science Foundation (NSF) has also seen a steady decline since the early 2010s (see Figure \ref{fig:budget}).


Furthermore, NASA's current budget is intended to provide funding for the research and operation of dozens of current and planned missions including the Hubble Space Telescope, James Webb Space Telescope, and the Nancy Grace Roman Space Telescope. Consequently, research grants and proposals have become considerably more competitive in recent years. Successful grant writing necessitates brevity as authors are required to summarize the background of their field and merit of their proposal in only a few pages. Additionally, studies in scientific communication have shown that unnecessary figures and tables can diminish the readability and clarity of a paper \citep{Bunkers23}. Smaller articles can even help develop readers' confidence and their understanding of the content \citep{ryba2021}. Some journals, such as AAS and Astronomy \& Astrophysics (A\&A), have very strict policies on the size and quanta of a paper to encourage short and concise papers, and judicial use of visual elements. Recently, \citet{aanda25} have introduced extra fees on longer papers. The implementation of \gfpm\ in articles and proposals thus reinforces the benefits of brevity while simultaneously producing interesting figures.

\section{Conclusion}
In this paper, we introduce the \gfpm, an inclusive version of \texttt{pastamarkers}. 
Like its precursors, \gfpm\ makes unique figures and visual elements, but the key characteristic for \gfpm\ lies in its inherent crumbling tendency that promotes smaller, clearer, and more concise scientific articles. 
This also helps authors manage publication costs, which tend to increase with the length of a paper. To the best of our knowledge, this work is the first to provide such a novel solution to the long-standing funding issues faced globally by the writing community. 
\bibliography{bibliography}{}
\bibliographystyle{aasjournalv7}

\end{document}